\documentclass[12pt]{article}
\usepackage{a4}
\usepackage{amsfonts}
\usepackage{amsmath,amssymb}

\def\un#1{\relax\ifmmode\@@underline#1\else
        $\@@underline{\hbox{#1}}$\relax\fi}

\let\du=\du                     

\def\j{\psi}

\def\p{\pi}
\def\q{\theta}

\def\s{\sigma}

\def\vq{\vartheta}

\def\ce{{\cal E}}

\def\car{{\cal R}}

\def\bo{{\raise-.3ex\hbox{\large$\Box$}}}               
\def\pa{\partial}                                       
\def\TH{{\raise.2ex\hbox{$\displaystyle \bigodot$}\mskip-4.7mu \llap H \;}}
\def\face{{\raise.2ex\hbox{$\displaystyle \bigodot$}\mskip-2.2mu \llap {$\ddot
        \smile$}}}                                      

\def\abs#1{\left| #1\right|}                    
\def\leftrightarrowfill{$\mathsurround=0pt \mathord\leftarrow \mkern-6mu
        \cleaders\hbox{$\mkern-2mu \mathord- \mkern-2mu$}\hfill
        \mkern-6mu \mathord\rightarrow$}
\def\dvec#1{\vbox{\ialign{##\crcr
        \leftrightarrowfill\crcr\noalign{\kern-1pt\nointerlineskip}
        $\hfil\displaystyle{#1}\hfil$\crcr}}}           

\def\frac#1#2{{\textstyle{#1\over\vphantom2\smash{\raise.20ex
        \hbox{$\scriptstyle{#2}$}}}}}                   
\def\sfrac#1#2{{\vphantom1\smash{\lower.5ex\hbox{\small$#1$}}\over
        \vphantom1\smash{\raise.4ex\hbox{\small$#2$}}}} 
\def\bfrac#1#2{{\vphantom1\smash{\lower.5ex\hbox{$#1$}}\over
        \vphantom1\smash{\raise.3ex\hbox{$#2$}}}}       
\def\afrac#1#2{{\vphantom1\smash{\lower.5ex\hbox{$#1$}}\over#2}}    

\def\[{\lfloor{\hskip 0.35pt}\!\!\!\lceil}
\def\]{\rfloor{\hskip 0.35pt}\!\!\!\rceil}
\def\Lag{{\cal L}}
\def\du#1#2{_{#1}{}^{#2}}

\def\ha{{\fracmm12}}

\def\un{\underline}
\def\fracmm#1#2{{{#1}\over{#2}}}

\def\low#1{{\raise -3pt\hbox{${\hskip 0.75pt}\!_{#1}$}}}

\newskip\humongous \humongous=0pt plus 1000pt minus 1000pt
\def\caja{\mathsurround=0pt}
\def\eqalign#1{\,\vcenter{\openup2\jot \caja
        \ialign{\strut \hfil$\displaystyle{##}$&$
        \displaystyle{{}##}$\hfil\crcr#1\crcr}}\,}
\newif\ifdtup

\newcommand{\be}{\begin{equation}}
\newcommand{\ee}{\end{equation}}
\newcommand{\nbe}{\begin{equation*}}
\newcommand{\nee}{\end{equation*}}

\newcommand{\lb}{\label}

\begin{document}

\thispagestyle{empty}

{\hbox to\hsize{
\vbox{\noindent October 2011 \hfill version 2 }}}

\noindent
\vskip2.0cm
\begin{center}

{\large\bf     COSMOLOGICAL CONSTANT in $F(\car)$ 
SUPERGRAVITY~\footnote{Supported in part by TMU Faculty of Science and
Engineering, and German SFB 676 }}
\vglue.3in

          Sergei V. Ketov~${}^{a,b,c}$ and Natsuki Watanabe~${}^a$
\vglue.1in

${}^a$ {\it Department of Physics, Graduate School of Science, 
   Tokyo Metropolitan University, Hachioji-shi, Tokyo 192-0397, Japan}\\
${}^b$ {\it Institute for Physics and Mathematics of Universe, The 
           University of Tokyo, Kashiwa-shi, Chiba 277-8568, Japan}\\
${}^c$ {\it DESY Theory Group, Notkestrasse 85, 22607 Hamburg, Germany}\\
\vglue.1in
ketov@phys.se.tmu.ac.jp, watanabe-natsuki1@ed.tmu.ac.jp
\end{center}

\vglue.3in

\begin{center}
{\Large\bf Abstract}
\end{center}
\vglue.1in

\noindent A cosmological constant in the regime of low spacetime curvature is calculated
in the recently proposed version of $F(\car)$ supergravity with a generic cubic function $F$.
The $F(\car)$ supergravity is the $N=1$ supersymmetric extension of $f(R)$ gravity. The cubic 
model is known to successfully describe a chaotic (slow-roll) inflation in the regime of high 
spacetime curvature. We find that a simple extension of the same model allows a {\it positive} 
cosmological constant in the regime of low spacetime curvature. The inflaton superfield in 
$F(R)$ supergravity (like inflaton in $f(R)$ gravity) violates the Strong Energy Condition and 
thus breaks the restriction of the standard supergravity (with usual matter) that can only have 
either a negative or vanishing cosmological constant.  

\newpage

\section{Introduction}

The Standard ($\Lambda$-CDM) Model in cosmology gives a phenomenological description of
the observed {\it Dark Energy} (DE) and {\it Dark Matter} (DM). It is based on the
use of a small positive cosmological constant $\Lambda$ and a {\it Cold Dark Matter} (CDM), 
and is consistent with all observations coming from the existing cosmological, Solar
system and ground-based laboratory data. However, the $\Lambda$-CDM  Model cannot be the 
ultimate answer to DE, since it implies its time-independence. For example, the 
`primordial' DE responsible for inflation in the early universe was different from $\Lambda$ 
and unstable.  The {\it dynamical}  (ie. time-dependent) models of DE can be easily 
constructed by using the $f(R)$ gravity theories, defined via replacing the scalar 
curvature $R$ by a function $f(R)$ in the gravitational action. The $f(R)$ gravity 
provides the self-consistent non-trivial alternative to the $\Lambda$-CDM Model --- see 
eg., refs.~\cite{stu,fr,odin} for a review. The use of $f(R)$ gravity in the 
inflationary cosmology was pioneered by Starobinsky \cite{star}. Viable 
$f(R)$-gravity-based models of the current DE are also known \cite{cde1,cde2,cde3}, and 
the combined inflationary-DE models are possible too \cite{stu}. 

Despite of the apparent presence of the higher derivatives, an $f(R)$ gravity theory 
can be free of ghosts and tachyons. The corresponding stability conditions are well 
known -- see Sec.~2 below. Under those conditions, it is always possible to prove the
classical equivalence of an $f(R)$ gravity theory to the certain scalar-tensor 
theory of gravity \cite{eq,bac,ma}. Dynamics of the spin-2 part of metric in $f(R)$
gravity (compared to Einstein gravity) is not modified, but there is the extra 
propagating scalar field (called scalaron) given by the spin-0 part of metric. By the 
classical equivalence above we mean that both theories lead to the same inflaton scalar 
potential and, therefore, the same inflationary dynamics. However, the physical nature 
of inflaton in each theory is different. In the $f(R)$ gravity and $F(\car)$ supergravity
inflaton field is part of metric, whereas in the scalar-tensor gravity and supergravity 
inflaton is a matter particle. Therefore, the inflaton interactions with other matter fields 
are different in both theories. It gives rise to different inflaton decay rates and different 
reheating in the post-inflationary universe. 

In our recent papers \cite{our1,our2,our3,our4,our5,our6,our7,our8} we proposed the new 
supergravity theory (we call it $F(\car)$-supergravity), and studied some of its 
physical applications (see also refs.~\cite{our9,our10} for our earlier related work).
The $F(\car)$-supergravity can be considered as the $N=1$ locally supersymmetric 
extension of $f(R)$ gravity in four space-time dimensions.~\footnote{Another 
(unimodular) $F(R)$ supergravity theory was proposed in ref.~\cite{nish}.} Supergravity 
is well-motivated in High-Energy Physics theory beyond the Standard Model of 
elementary particles. Supergravity is also the low-energy effective action of 
Superstrings. As was demonstrated in ref.~\cite{our1}, an $F(\car)$ supergravity is 
classically equivalent to the $N=1$ Poincar\'e supergravity coupled to a dynamical 
(quintessence) chiral superfield, whose (non-trivial) K\"ahler potential and superpotential 
are dictated by the chiral (holomorphic) function $F$. The classical equivalence is achieved 
via a non-trivial field redefinition \cite{our1} that gives rise to a non-trivial Jacobian in
the path integral formulation of those quantum field theories (below their unitarity bounds).
Hence, their classical equivalence is expected to be broken in quantum theory.~\footnote{See
ref.~\cite{kmy} for the first steps in quantizing $f(R)$ gravity theories.}

The natural embedding of the Starobinsky $(R+R^2)$-inflationary model into $F(\car)$ 
supergravity was found in ref.~\cite{our7}. It provides the very economical realization 
of chaotic inflation (at early times) in supergravity, which is consistent with observations 
\cite{wmap7} and gives a simple solution to the $\eta$-problem in supergravity \cite{eta}.

The natural question arises, whether $F(\car)$ supergravity is also capable to describe
the present DE or have a positive cosmological constant. It is non-trivial because the 
standard  supergravity with usual matter can only have a negative or vanishing cosmological 
constant \cite{gibb}. It takes place since the usual (known) matter does not violate the 
{\it Strong Energy Condition} (SEC) \cite{hawe}. A violation of SEC is required for an 
accelerating universe, and it is easily achieved in $f(R)$ gravity due to the fact that the
quintessence field in $f(R)$ gravity is part of metric (ie. the unusual matter). Similarly, 
the quintessence scalar superfield in $F(\car)$ supergravity is part of super-vielbein, and 
it also gives rise to a violation of SEC. In  this Letter we further extend the Ansatz used 
in ref.~\cite{our7} for $F$-function, and apply it to get a positive cosmological constant 
in the regime of low spacetime curvature (at late times).

Our paper is organized as follows. In sec.~2 we briefly recall the superspace 
construction of $F(\car)$ supergravity, its relation to $f(R)$ gravity and the stability
conditions. In sec.~3 we define our model of $F(\car)$ supergravity, and compute its
cosmological constant. Sec.~4 is our conclusion. 

Throughout the paper we use the units $c=\hbar=M_{\rm Pl}=1$ in terms of the (reduced) 
Planck mass $M_{\rm Pl}$, with the spacetime signature $(+,-,-,-)$. Our basic notation 
of General Relativity coincides with that of ref.~\cite{landau}. An AdS-spacetime has a 
positive scalar curvature, and a dS-spacetime has a negative scalar curvature in our 
notation.

\section{$F(\car)$ supergravity and $f(R)$ gravity}

A concise and manifestly supersymmetric description of supergravity is given
by superspace. We refer the reader to the textbooks \cite{ss1,ss2,ss3} for details of 
the superspace formulation of supergravity. A construction of the $F(\car)$ 
supergravity action goes beyond the supergravity textbooks.

The most succinct formulation of $F(\car)$ supergravity exist in a chiral
4D, $N=1$ superspace where  it is defined by the action \cite{our1}
\be  \lb{act}
 S_F = \int d^4x d^2\q\, \ce F(\car) + {\rm H.c.}
\ee
in terms of a holomorphic function $F(\car)$ of the covariantly-chiral scalar
curvature superfield $\car$, and the chiral superspace density $\ce$. The chiral
$N=1$ superfield $\car$ has the scalar curvature $R$ as the field coefficient
at its $\q^2$-term. The chiral superspace density $\ce$ (in a WZ gauge) reads
\be \lb{cde}
\ce = e \left( 1- 2i\q\s_a\bar{\j}^a +3\q^2 X\right) 
\ee
where $e=\sqrt{-g}$, $\j^a$ is gravitino, and $X=S-iP$ is the complex scalar 
auxiliary field (it does not propagate in the theory (\ref{act}) despite of the
apparent presence of the higher derivatives \cite{our1}). 

A bosonic $f(R)$ gravity action is given by \cite{stu,fr,odin}
\be \lb{mgrav}
 S_f = \int d^4x \,\sqrt{-g}\, f(R) \ee
in terms of the real function $f(R)$ of the scalar curvature $R$. The relation between 
the master chiral superfield  function $F(\car)$ in eq.~(\ref{act}) and the 
corresponding bosonic function $f(R)$ in eq.~(\ref{mgrav}) can be established by 
applying the standard formulae of superspace \cite{ss1,ss2,ss3} and ignoring the 
fermionic contributions. For simplicity, we also ignore the complex nature of $F$ and 
$X$ in what follows.

The embedding of $f(R)$ gravity into $F(\car)$ supergravity is given by 
\cite{our1,our2,our3}
\be \lb{emb1}
f(R) = f(R,X(R))
\ee
where the function $f(R,X)$ (or the gravity Lagrangian $\Lag$) is defined by
\be \lb{emb2}
\Lag= f(R,X) = 2F'(X) \left[ \fracmm{1}{3} R +4 X^2 \right] + 6XF(X)   
\ee
and the function $X=X(R)$ is determined by solving an algebraic equation,
\be \lb{emb3}
 \fracmm{\pa f(R,X)}{\pa X} =0 
\ee
The primes denote the derivatives with respect to the given argument. Equation 
(\ref{emb3}) arises by varying the action (\ref{act}) with respect to the 
auxiliary field $X$. It cannot be explicitly solved for $X$ in a generic $F(\car)$ 
supergravity theory.

The cosmological constant in $F(\car)$ supergravity, in the regime of low space-time 
curvature, is thus given by
\be \lb{cc1}
\Lambda = - f(0,X_0)  
\ee
where $X_0=X(0)$. It should be mentioned that $X_0$ represents the vacuum expectation
value of the auxiliary field $X$ that determines the scale of the supersymmetry 
breaking. Both inflation and DE imply $X_0\neq 0$.

The $f(R)$-gravity stability conditions in our notation are given by \cite{stu,our6}
\be \lb{csta} 
f'(R) < 0 
\ee
and
\be \lb{qsta} 
 f''(R) > 0
\ee
The first (classical stability) condition (\ref{csta}) is related to the sign factor 
in front of the Einstein-Hilbert term (linear in $R$) in the $f(R)$-gravity action, 
and it ensures that graviton is not a ghost. The second (quantum stability) condition 
(\ref{qsta}) ensures that scalaron is not a tachyon. In $F(R)$ supergravity
eq.~(\ref{csta}) is replaced by a stronger condition \cite{our6},
\be \lb{cs}
F'(X) < 0 
\ee
Equation (\ref{cs}) guarantees the classical stability of the $f(R)$-gravity embedding 
into the full $F(\car)$ supergravity against small fluctuations of the axion field $P$
\cite{our6}.

To describe the early universe inflation (ie. in the regime of {\it high} spacetime 
curvature $R\to -\infty$), the function $f(R)$ should have the profile
\be \lb{infp} 
f(R) = - \fracmm {1}{2}R + R^2 A(R) \equiv f_{\rm EH}(R) +  R^2 A(R)
\ee
with the slowly varying function $A(R)$ in the sense
\be \lb{svar}
 \abs{A'(R)} \ll \abs{ \fracmm{A(R)}{R} } \quad {\rm and}
 \abs{A''(R)} \ll \abs{ \fracmm{A(R)}{R^2} } 
\ee
The simplest choice $A=const. >0$ gives rise to the Starobinsky model \cite{star} 
with
\be \lb{star1}
f_S(R) = -\fracmm{1}{2}R +\fracmm{R^2}{12M^2_{\rm inf}}
\ee
where the inflaton (scalaron) mass $M_{\rm inf}$ has been introduced.

To describe DE in the present universe, ie. in the regime with {\it low} spacetime 
curvature $R$, the function $f(R)$ should be close to the Einstein-Hilbert (linear) 
function $f_{\rm EH}(R)$ with a small positive $\Lambda$, 
\be \lb{dep}
\abs{f(R)-f_{\rm EH}(R) } \ll \abs{f_{\rm EH}(R)},\quad 
\abs{f'(R)-f'_{\rm EH} } \ll 1,\quad
\abs{Rf''(R)} \ll 1  
\ee   
ie. $f(R)\approx -\fracmm{1}{2}R -\Lambda$ for small $R$ with the very small and positive 
$\Lambda\approx 10^{-118}(M^4_{\rm Pl})$.

\section{Cosmological constant}

Equations (\ref{emb2}) and (\ref{cc1}) imply
\be \lb{cc2} 
\Lambda =-8F'(X_0)X_0^2 -6X_0F(X_0) 
\ee
where $X_0$ is a solution to the algebraic equation
\be \lb{cc3}
4X_0^2F''(X_0) + 11X_0F'(X_0) + 3F(X_0) =0
\ee
As is clear from eq.~(\ref{cc2}), to have $\Lambda\neq 0$, one must have $X_0\neq 0$, ie.
a (spontaneous) supersymmetry breaking. However, in order to proceed further, one
needs a reasonable Ansatz for the $F$-function in eq.~(\ref{act}).

The simplest opportunity is given by expanding the function $F(\car)$ in Taylor 
series with respect to $\car$. Since the $N=1$ chiral superfield $\car$ has
$X$ as its leading field component (in $\q$-expansion), one may expect that the Taylor
expansion is a good approximation as long as $\abs{X_0}\ll 1(M_{\rm Pl})$. As was
demonstrated in ref.~\cite{our7}, a viable (successful) description of inflation is
possible in $F(\car)$ supergravity, when keeping the {\it cubic} term $\car^3$ in the
Taylor expansion of the $F(\car)$ function. It is, therefore, natural to expand the
function $F$ up to the cubic term with respect to $\car$, and use it as our Ansatz 
here,
\be \lb{an}
 F(\car) = f_0  - \fracmm{1}{2}f_1\car + \fracmm{1}{2}f_2\car^2 
-\fracmm{1}{6}f_3\car^3 
\ee
with some real coeffieints $f_0,f_1,f_2,f_3$. The Ansatz (\ref{an}) differs from the
one used in ref.~\cite{our7} by the presence of the new parameter $f_0$ only. It is
worth emphasizing here that $f_0$ is {\it not} a cosmological constant because one 
still has to eliminate the auxiliary field $X$. The stability conditions in the case
(\ref{an}) require
\be \lb{sa1}
f_1>0~~, \qquad f_2>0~~,\qquad f_3 > 0 
\ee
and
\be \lb{sa2}
f_2^2 < f_1 f_3
\ee
Inflation requires $f_3\gg 1$ and $f_2^2\gg f_1$ \cite{our7}.~\footnote{The stronger condition 
$f^2_2\ll f_1f_3$ was used in ref.~\cite{our7} for simplicity.} 
As was shown in ref.~\cite{our7}, in the high-curvature regime the
effective $f(R)$-gravity action (originating from the $F(\car)$ supergravity defined
by eqs.~(\ref{act}) and (\ref{an}) with $f_0=0$) takes the form of eq.~(\ref{star1}) 
with $f_3=15M^2_{\rm inf}$. To meet the WMAP observations \cite{wmap7}, the parameter
$f_3$ should be approximately $6.5\cdot 10^{10}(N_e/50)^2$, where $N_e$ is the number 
of e-foldings \cite{our7}. The cosmological constant in the high-curvature regime 
does not play a significant role and may be ignored there.

In the low curvature regime, in order to recover the Einstein-Hilbert term, one has
to fix $f_1=3/2$ \cite{our7}. Then the Ansatz (\ref{an}) leads to the gravitational Lagrangian 
\be 
\lb{frx}
f(R,X) = -5f_3X^4 +11f_2X^3 - \fracmm{1}{3}f_3\left(R+\fracmm{63}{2f_3}\right)
X^2 + \left(6f_0 + \fracmm{2}{3}f_2R\right)X - \fracmm{1}{2}R 
\ee
and the auxiliary field equation
\be \lb{afe}
X^3 - \fracmm{33f_2}{20f_3}X^2 +\fracmm{1}{30}\left( R + \fracmm{63}{2f_3}\right)
X - \fracmm{1}{30f_3}\left( f_2R +9f_0\right)=0
\ee
whose formal solution is available via the standard Cardano (Vi\`ete) 
formulae \cite{cv}.

In the low-curvature regime we find a cubic equation for $X_0$ in the form
\be \lb{cub}
 X_0^3 - \left(\fracmm{33f_2}{20f_3}\right)X_0^2 
+\left(\fracmm{21}{20f_3}\right)X_0 -\left(\fracmm{3f_0}{10f_3}\right)=0
\ee

`Linearizing' eq.~(\ref{cub}) with respect to $X_0$ brings the solution $X_0=2f_0/7$
whose substitution into the action (\ref{frx}) gives rise to a {\it negative} cosmological 
constant, $\Lambda_0=-6f_0^2/7$. This way we recover the standard supergravity case.

Equations (\ref{frx}) and (\ref{cub}) allow us to write down the exact eq.~(\ref{cc1})
for the cosmological constant in the factorized form
\be \lb{cce}
\Lambda(X_0) = - \fracmm{11f_2}{4}X_0 ( X_0-X_-)(X_0-X_+)
\ee
where $X_{\pm}$ are the roots of the quadratic equation $x^2-\fracmm{21}{11f_2}x
+\fracmm{18f_0}{11f_2}=0$, ie.
\be \lb{roots}
X_{\pm} = \fracmm{21}{22f_2}\left[ 1\pm \sqrt{ 1- \fracmm{2^3\cdot 11}{7^2}f_0f_2}\right]
\ee
Since $f_0f_2$ is supposed to be very small, both roots  $X_{\pm}$ are real and positive.

Equation (\ref{cce}) implies that $\Lambda>0$ when either (I) $X_0<0$, or (II)  $X_0$ is inside
the interval $(X_-,X_+)$.

By using {\it Matematica} we were able to numerically confirm the existence of solutions to
eq.~(\ref{cub}) in the region (I) when $f_0<0$, but not in the region (II). So, to this end, 
we continue with the region (I) only.  All real roots of eq.~(\ref{cub}) are given by
\be \lb{3roots}
\eqalign{
(X_0)_1 = ~&~ 2\sqrt{-Q} \cos\left(\fracmm{\vq}{3}\right) +\fracmm{11f_2}{20f_3}~~,\cr
(X_0)_2 = ~&~ 2\sqrt{-Q} \cos\left(\fracmm{\vq+2\p}{3}\right) +\fracmm{11f_2}{20f_3}~~,\cr
(X_0)_3 = ~&~ 2\sqrt{-Q} \cos\left(\fracmm{\vq+4\p}{3}\right) +\fracmm{11f_2}{20f_3}~~,\cr }
\ee
in terms of the Cardano-Vi\`ete parameters
\be \lb{cvp} 
\eqalign{
Q= ~&~ -\fracmm{11f_2}{2^2\cdot 5f_3} -\fracmm{7^2}{2^4\cdot5^2 f_3^2}\approx -\fracmm{11f_2}{20f_3}~~,\cr
R= ~&~ -\fracmm{3\cdot 7\cdot 11f_2}{2^5\cdot 5^2 f^2_3} + \fracmm{3f_0}{2^2\cdot 5f_3}
+\fracmm{11^3f_2^3}{2^6\cdot 5^3f_3^3} \approx -\fracmm{1}{20f_3}\left(-\fracmm{21}{2}Q+3f_0\right)~~\cr}
\ee
and the angle $\vq$ defined by  
\be \lb{vq}
\cos\vq =\fracmm {R}{\sqrt{-Q^3}}
\ee
The Cardano discriminant reads $D=R^2+Q^3$. All three roots are real provided that $D<0$. It is
known to be the case in the high-curvature regime \cite{our7}, and it is also the case when $f_0$ 
is extremely small. Under our requirements on the parameters the angle $\vq$ is very close to zero, 
so the relevant solutions $X_0<0$ are given by the 2nd and 3rd lines of eq.~(\ref{3roots}), with 
$X_0\approx f_0/10$. 

\section{Conclusion}

We demonstrated that it is possible to have a {\it positive} cosmological constant (at low spacetime 
curvature or late times) in the particular $F(\car)$ supergravity (without its coupling to 
super-matter) described by the Ansatz (\ref{an}). The same Ansatz is applicable for describing 
a viable chaotic inflation in supergravity (at high spacetime curvature or early times). The positive
 cosmological constant was technically achieved as the {\it non-linear} effect with respect to the 
superspace curvature and  spacetime curvature in the relatively narrow part of the parameter space 
(it is, therefore, highly constrained). 

In the particular $F(\car)$ supergravity model we considered, the effective $f(R)$ gravity function 
is essentially given by the Starobinsky function $(-\ha R+\fracmm{1}{12M^2_{\rm inf}} R^2)$ in the 
high curvature regime, and by the DE-like function $(-\ha R-\Lambda)$ in the low curvature 
regime. Therefore, our model has a cosmological solution which describes an inflationary universe 
of the quasi-dS type with the Hubble function $H(t)\approx \fracmm{M^2_{inf}}{6}(t_{end}-t)$ at early
 times $t< t_{end}$ and an accelerating universe of the dS-type with $H=\Lambda$ at late times. 
It is similar to the known cosmological solutions unifying inflation and DE in $f(R)$ gravity 
\cite{stu}. 

Of course, describing the DE in the present universe requires an enormous fine-tuning 
of our parameters in the $F$-function. However, it is the common feature of all known 
approaches to the DE. This paper does not contribute to `explaining' the smallness of 
the cosmological constant.

\end{document}